%% file: main.tex
\documentclass[10pt, aps,prc,twocolumn,superscriptaddress,preprintnumbers,
amsmath, 
floatfix,
longbibliography,
nofootinbib
]{revtex4-1}
\usepackage[T1]{fontenc}
\usepackage[utf8x]{inputenc} 
\usepackage{adjustbox}          
\usepackage[caption=false]{subfig}
\usepackage{url}
\usepackage{color}
\usepackage{float}
\usepackage[pdftex,colorlinks=true, linkcolor = blue, citecolor=blue,urlcolor=blue, bookmarksnumbered=true, bookmarksopen=true]{hyperref}
\usepackage{longtable}
\usepackage{amsfonts}
\usepackage{dsfont}
\usepackage{wrapfig,bm} 
\usepackage[normalem]{ulem}
\usepackage{MnSymbol}

\newcommand{\beq}{\begin{equation}}
\newcommand{\eeq}{\end{equation}}
\newcommand{\bea}{\begin{eqnarray}}
\newcommand{\eea}{\end{eqnarray}}

\begin{document}
\title{Influence of the Exit Channel in $^{235}$U(n,f) and $^{239}$Pu(n,f) Reactions in Time-Dependent Density Functional Theory}

  \author{Ibrahim Abdurrahman}
\affiliation{ Facility for Rare Isotope Beams, Michigan State University, East Lansing, Michigan 48824, USA}
 \affiliation{Department of Physics and Astronomy, Michigan State University, East Lansing, Michigan 48824, USA}
 \affiliation{ Theoretical Division, Los Alamos National Laboratory, Los Alamos, New Mexico 87545, USA}

 \author{Matthew Kafker}
 \affiliation{Cyclotron Institute, Texas A\&M University, College Station, Texas 77843, USA}
\affiliation{Department of Physics,%
  University of Washington, Seattle, Washington 98195--1560, USA} 

 \author{Aurel Bulgac }
\affiliation{Department of Physics,%
  University of Washington, Seattle, Washington 98195--1560, USA} 

  \author{Ionel Stetcu}
 \affiliation{ Theoretical Division, Los Alamos National Laboratory, Los Alamos, New Mexico 87545, USA}
 
\date{\today}

\begin{abstract}
This study investigates the consequences of the intrinsic deformation of the fissioning nuclear system near the outer saddle point on the shape evolution of the nucleus from saddle to scission and on the properties of the fission fragments. It is found that trajectories generally split into at least three classes, asymmetric, near-symmetric, and highly-asymmetric fission, that are roughly determined by the initial magnitude of the octupole moment, and each of which exhibits different scission dynamics and fragment properties. Near-symmetric modes result in a highly elongated neck at scission, leading to the neck rupture occurring when the proto-fragments are further apart than is the case for typical asymmetric fission, which in turn leads to a lower total kinetic energy and higher total excitation energy.  The majority of this additional excitation energy goes into the heavy fission fragment, that develops a substantial quadrupole deformation. A similar trend is observed for the total kinetic energy of highly-asymmetric fission events, and the opposite trend for the excitation energy of the fission fragments as the majority of the additional excitation energy goes into the light fission fragment instead. The study also characterizes the neck rupture, including its effect on the emission of scission neutrons in near-symmetric fission. 
\end{abstract} 

\preprint{NT@UW-26-16}
\preprint{LA-UR-26-25651}

\maketitle


\section{Introduction}\label{sec:intro}

In 1939, nuclear fission was discovered experimentally by Hahn and Strassmann~\cite{Hahn:1939}, and shortly thereafter its main mechanism was explained by Meitner and Frisch~\cite{Meitner:1939}.  Close to a century later, fission still lacks a complete microscopic description, in part because it is a highly complex quantum many-body process with qualitatively distinct stages, each of which occurring at vastly different timescales~\cite{Gonnenwein:2014}.  This study focuses on the most rapid and non-equilibrium stage of fission, the descent of the fissioning nuclear system (FNS) from the outer turning point to scission.\footnote{Note that the system is described by a new term, FNS, instead of the previously used term, compound nucleus (CN).  This distinction is necessary, as the nucleus is approximated by a generalized Slater Determinant, which is not of sufficient complexity to correspond to N. Bohr's CN~\cite{Bohr:1936}. In his description, where the idea of a CN was first established, a CN wave function corresponds to a very long-lived state in the continuum, arising from a superposition of a very large number $\geq {\cal O}(10^{4})$ of mean field configurations.} 
It is at this stage, which cannot be probed directly by experiment, that the properties of the primary fission fragments (FFs) are defined.  Using time-dependent density functional theory (TDDFT) extended to superfluid systems, known as the superfluid local density approximation (TDSLDA)~\cite{Shi:2021}, we investigate how the initial intrinsic deformation of the FNS  at the outer saddle impacts the emerging FF properties.  

For $^{235}$U(n$_{\mathrm{th}}$,f) and $^{239}$Pu(n$_{\mathrm{th}}$,f) reactions, the most prevalent fission mode is asymmetric fission, which corresponds to FNS s near the outer fission barrier with only a moderate octupole deformation.  This mode has been investigated extensively within TDSLDA~\cite{Bulgac:2016,Bulgac:2019c,Bulgac:2021,Bulgac:2022b,Scamps:2023a,Abdurrahman:2024,Bulgac:2025} and other variants of TDDFT~\cite{Ren:2022,Ren:2022a,Zhang:2023,Zhao:2022,Zhao:2023}.\footnote{In these other variants, all TDDFT simulations start from initial states energy at 1 MeV below the ground state of the FNS with a correspondingly large quadrupole and octupole deformation. This energy is well below the energy of the outer saddle barrier: $\approx 7-8$ MeV below. As a result, the role of the dissipative fission dynamics in the descent from the outer fission barrier, up to their initial configurations, is neglected. Furthermore, in these studies the authors use the Time-Dependent Bardeen-Cooper-Schrieffer approximation to describe the role of pairing correlations, which is known to violate the one body continuity equation~\cite{Scamps:2012}. } In contrast, near-symmetric fission, where has a small octupole deformation at the outer saddle, and highly-asymmetric fission, where the FNS  has a large octupole deformation at the outer saddle, have only been addressed sparingly~\cite{Bulgac:2019c,Bulgac:2025,Bjelvcic:2026}. This is partly because, currently, there is a microscopic ambiguity for assigning a probability distribution to initial states at the outer saddle~\cite{Sadhukhan:2016,Sadhukhan:2017, Verriere:2020}, and as such typically only the most likely channel is treated. The microscopic ambiguity could potentially be resolved within an extension of the TDSLDA 
to the generator coordinate method that was recently suggested~\cite{Bulgac:2024e} as well as applied to multi-nucleon transfer reactions~\cite{Kafker:2026}. 
Highly asymmetric fission modes have been conjectured by \textcite{Brosa:1990} and studied by various authors in phenomenological models, see Refs.~\cite{Capote:2016,Capote:2016a,Ishizuka:2017,Ivanyuk:2024} and references therein.   

These modes are important to investigate microscopically, as they will become increasing populated at higher incident neutron energies. The neutron's energy will either effect the relative weights of the various configurations at the outer saddle, or, for fixed deformations, increase the temperature of the FNS. In prior microscopic investigations there is evidence that the initial temperature only weakly influences certain FF observables: FF spins, masses, charges, and the total kinetic energy (TKE)~\cite{Bulgac:2021,Bulgac:2025}. Consequently, the initial deformation of the FNS  at the saddle will likely be the most influential factor affecting the aforementioned observables as the initial excitation energy of the system increases.    

In this study several such fission modes are considered. All fission trajectories were performed on a $30\times30\times60$ cartesian point lattice with spacing $a=1$ fm using the LISE code~\cite{Shi:2021} with the SeaLL1~\cite{Bulgac:2018} energy density functional. The manuscript is organized in the following manner. In Sec.~\ref{sec:modes} the fission modes are precisely defined and the dynamics of the FNS  from saddle to scission is investigated. Sec.~\ref{sec:tke} focuses on the influence of these channels on the total kinetic energy (TKE) and FF excitation energies. Sec.~\ref{sec:neckr} investigates the neck rupture, including the emission of scission neutrons in near-symmetric fission. A summary of all key findings is provided in Sec.~\ref{sec:con}. 

\section{Neck Dynamics of Fission Modes}\label{sec:modes}

For $^{235}$U(n$_{\mathrm{th}}$,f) and $^{239}$Pu(n$_{\mathrm{th}}$,f) reactions, corresponding to FNS  $^{236}$U and $^{240}$Pu, three classes of saddle configurations are investigated: asymmetric, near-symmetric, and highly-asymmetric fission. Highly-asymmetric fission was only included for $^{236}$U. These classes are defined by the initial position of the FNS  on the potential energy surface (PES), which is characterized by the quadrupole and octupole moments,
\begin{equation}\label{eqn:deform}
\begin{split}
Q_{20} = \int (2 z'^2 - x'^2 - y'^2) n_t(\bm{r}) d^3r, \\
Q_{30} = \int z'  (2 z'^2 - 3 x'^2 - 3 y'^2) n_t(\bm{r}) d^3r,
\end{split}
\end{equation}
where $n_t(\bm{r})$ is the total number density, and $x' = x - x_{\mathrm{cm}}$, $y' = y - y_{\mathrm{cm}}$, and $z' = z - z_{\mathrm{cm}}$. 

\begin{figure} \includegraphics[width=1.0\columnwidth]{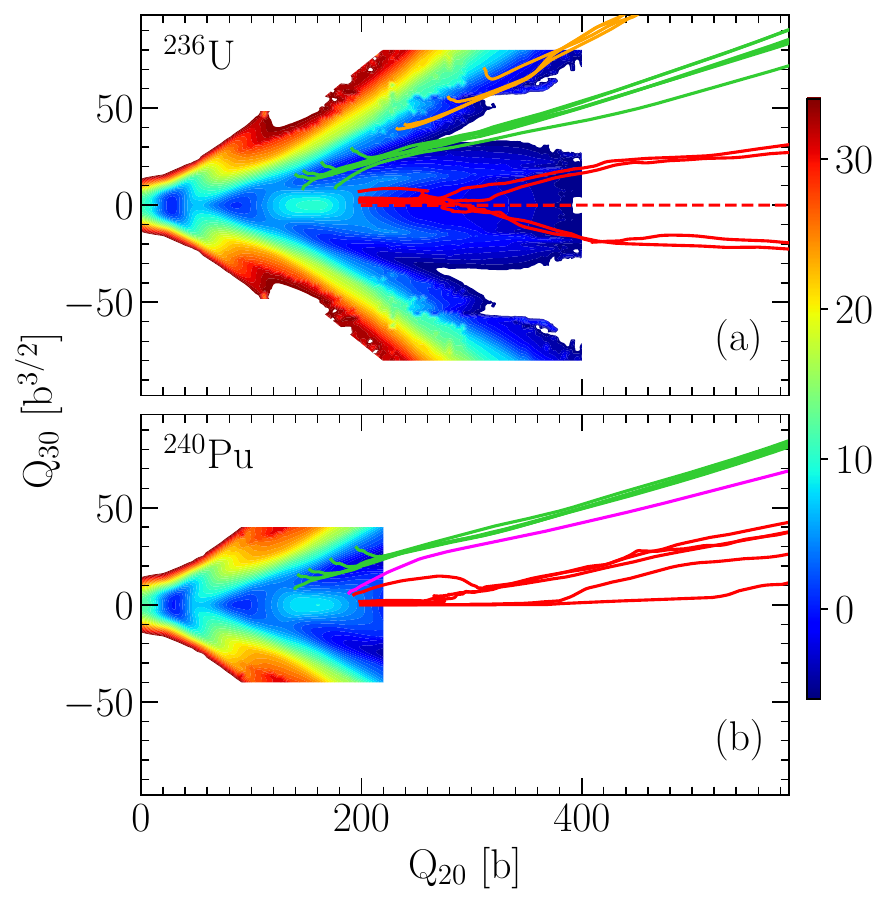}  \caption{ \label{fig:trajs}  Panels (a) and (b) show the potential energy surfaces (PES) for $^{236}$U and $^{240}$Pu respectively. The colors represent the minimum energy of the compound, at fixed octupole and quadrupole deformations, with respect to the ground state in MeV.  The lines represent fission trajectories, prepared at the outer saddle of the PES, which are evolved using the TDSLDA equations. Asymmetric trajectories are represented by solid green lines, near-symmetric trajectories are represented by solid red lines, and highly-asymmetric trajectories are represented by solid orange lines. In panel (a), one exactly symmetric trajectory is represented by a dashed red line.  In panel (b), the magenta line represents an ``intermediate'' fission trajectory, which begins close to the near-symmetric initial conditions, but  ultimately evolves toward the asymmetric valley on the PES.}  \end{figure}

In Fig.~\ref{fig:trajs} the various classes are identified by the initial positions of the fission trajectories on the PES. Asymmetric modes start on the ridge of the asymmetric fission valley, with $Q_{30} < 35 $ b$^{3/2}$. If $Q_{30}$ is approximately $ > 35$ b$^{3/2}$, the trajectories are classified as highly-asymmetric. Near-symmetric modes start on the ridge of the symmetric fission valley. More details are provided in the caption of Fig.~\ref{fig:trajs}.   

During the initial stage of the evolution a few interesting trends are observed. First, the mass asymmetry of the FNS  can flip during the evolution. This was also seen previously for $^{240}$Pu~\cite{Bulgac:2019c,Bjelvcic:2026} and for odd and odd-odd nuclei~\cite{Bulgac:2025}. Second, exactly symmetric fission is extremely unstable with respect to small perturbations of the compound's initial octupole deformation. For $^{240}$Pu two trajectories start with $Q_{30} \sim 10^{-7}$ b$^{3/2}$. Neither lead to symmetric fission. Instead the final mass asymmetry of the fragments is $A_L/A_H \sim 0.9$. For $^{236}$U one trajectory with initial $Q_{30} \sim 10^{-8}$ b$^{3/2}$ does successfully fission symmetrically. This explains why exactly symmetric fission is highly unlikely for fission induced by thermal neutrons, consistent with what is known experimentally. Third, there is a bifurcation at the transition between asymmetric and near-symmetric modes, where the trajectory will either evolve towards the asymmetric valley, shown by the magenta line in Fig.~\ref{fig:trajs}, or the symmetric valley. This suggests a forbidden zone (or more likely multiple forbidden zones) on the PES, at least in the absence of triaxial deformations or fluctuations~\cite{Bulgac:2019d}. These zones act to separate the various fission modes. This is also observed in~\cite{Bjelvcic:2026}, where considerably more trajectories were considered.  

\begin{figure} \includegraphics[width=1.0\columnwidth]{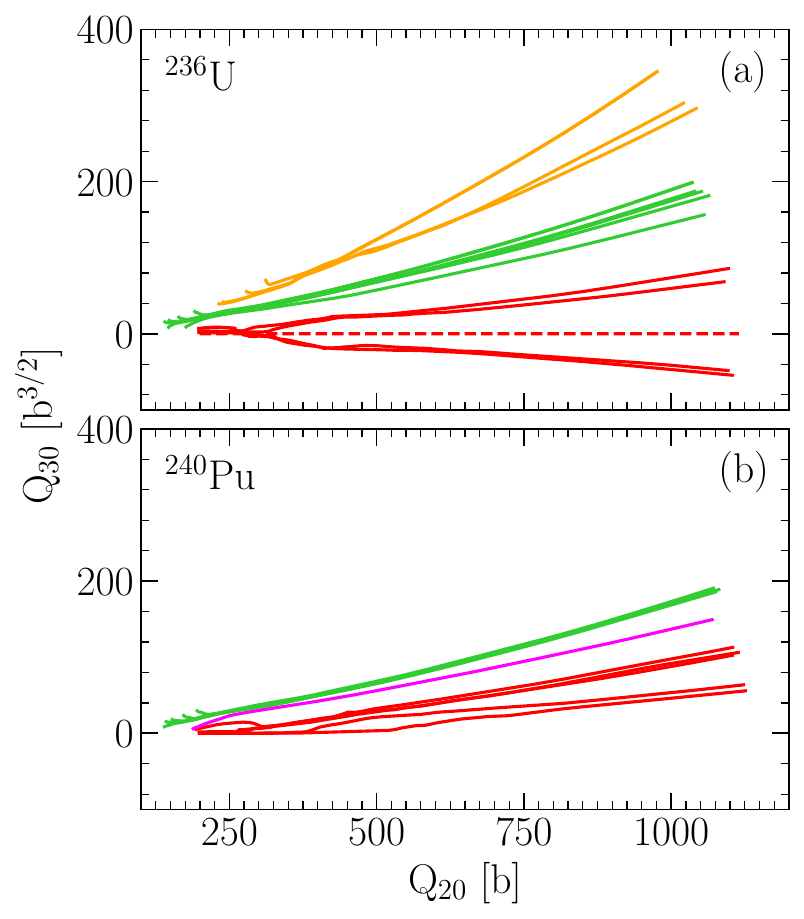}  \caption{ \label{fig:ltrajs}  Panels (a) and (b) show the evolution of the FNS  octupole and quadrupole deformation moments for $^{236}$U and $^{240}$Pu fission trajectories respectively. The colors represent the same fission modes described in the caption of Fig~\ref{fig:trajs}.}  \end{figure}

Fig.~\ref{fig:ltrajs} plots the entire evolution of the system from saddle to scission and beyond, up to when the center of masses of the FFs are separated by 30 fm. It shows that the clustering of the various fission modes are preserved throughout the evolution. The trajectories within each cluster spread as a function of time, which is reflective of the many possible final FF masses, as the FF masses are strongly correlated with the compound's final octupole deformation. The presence of highly-asymmetric fission trajectories also makes it clear that not all TDDFT trajectories starting along the outer ridge will evolve into the most likely fission pathway, which is often assumed when arguing that TDDFT cannot be used to obtain wide enough FF mass and charge distributions~\cite{Schunck:2022m}. This suggests there is at least a possibility that the inclusion of enough initial conditions in TDDFT will produce FF mass and charge distributions with comparable width to experiment and adiabatic TDGCM approaches~\cite{Verriere:2021}, even in the absence of fluctuations. It should be noted that although adiabatic methods obtain wide distributions~\cite{Verriere:2021}, comparable with experiment, they overestimate the population of symmetric and highly-asymmetric FFs, in addition to neglecting dissipation, which is known to be present in the saddle to scission stage of fission~\cite{Bulgac:2019c,Bender:2020}.

\begin{figure} \includegraphics[width=1.0\columnwidth]{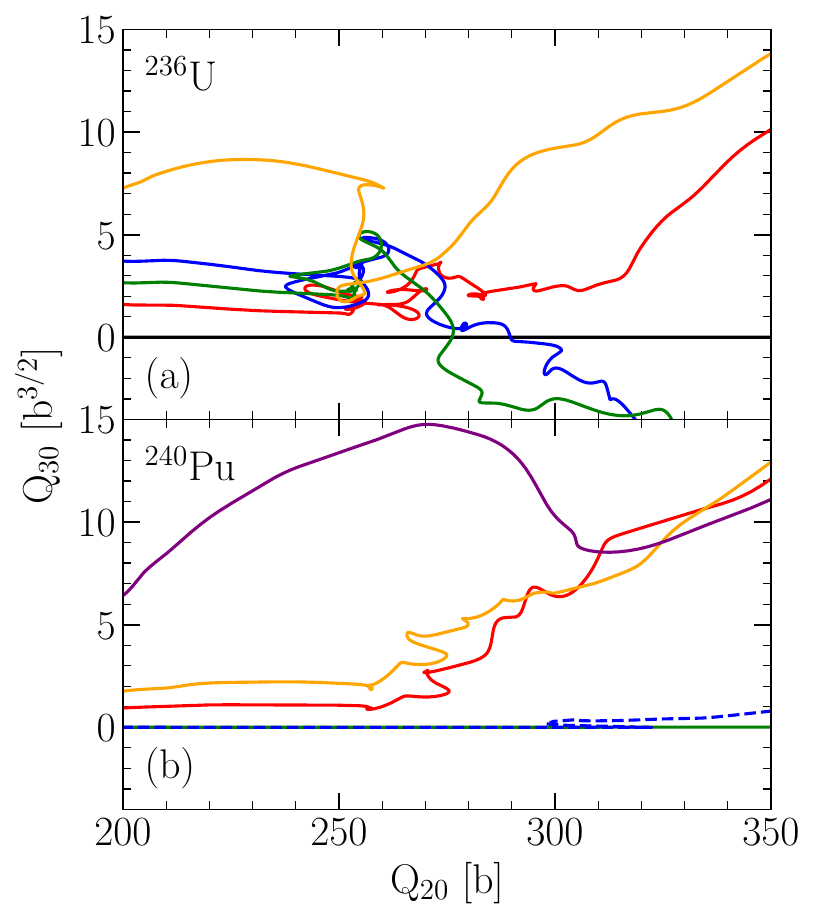}  \caption{ \label{fig:trajsym}  Panels (a) and (b) show the evolution of the FNS  octupole and quadrupole deformation moments for near-symmetric fission trajectories for $^{236}$U and $^{240}$Pu trajectories respectively.  The black line in the top panel represents an exactly symmetric fission trajectory. Both panels show the interval of deformations where the shape evolution is most convoluted.} \end{figure}

The dynamics of this stage can be highly non-trivial, especially for rarer fission modes. Fig.~\ref{fig:trajsym} zooms into the region of the PES, where the near-symmetric trajectories spend the majority of their time. The evolution is convoluted, as was observed in the odd-even FNS  $^{241}$Pu~\cite{Bulgac:2025}, although slightly less so. The time for these trajectories to evolve from the outer fission barrier to scission is significantly increased. No ``convoluted trajectories'' were seen for asymmetric fission, unlike what was established in the TDSLDA treatment of odd and odd-odd nuclei induced fission~\cite{Bulgac:2025}. 

\begin{figure*} \includegraphics[width=2.0\columnwidth]{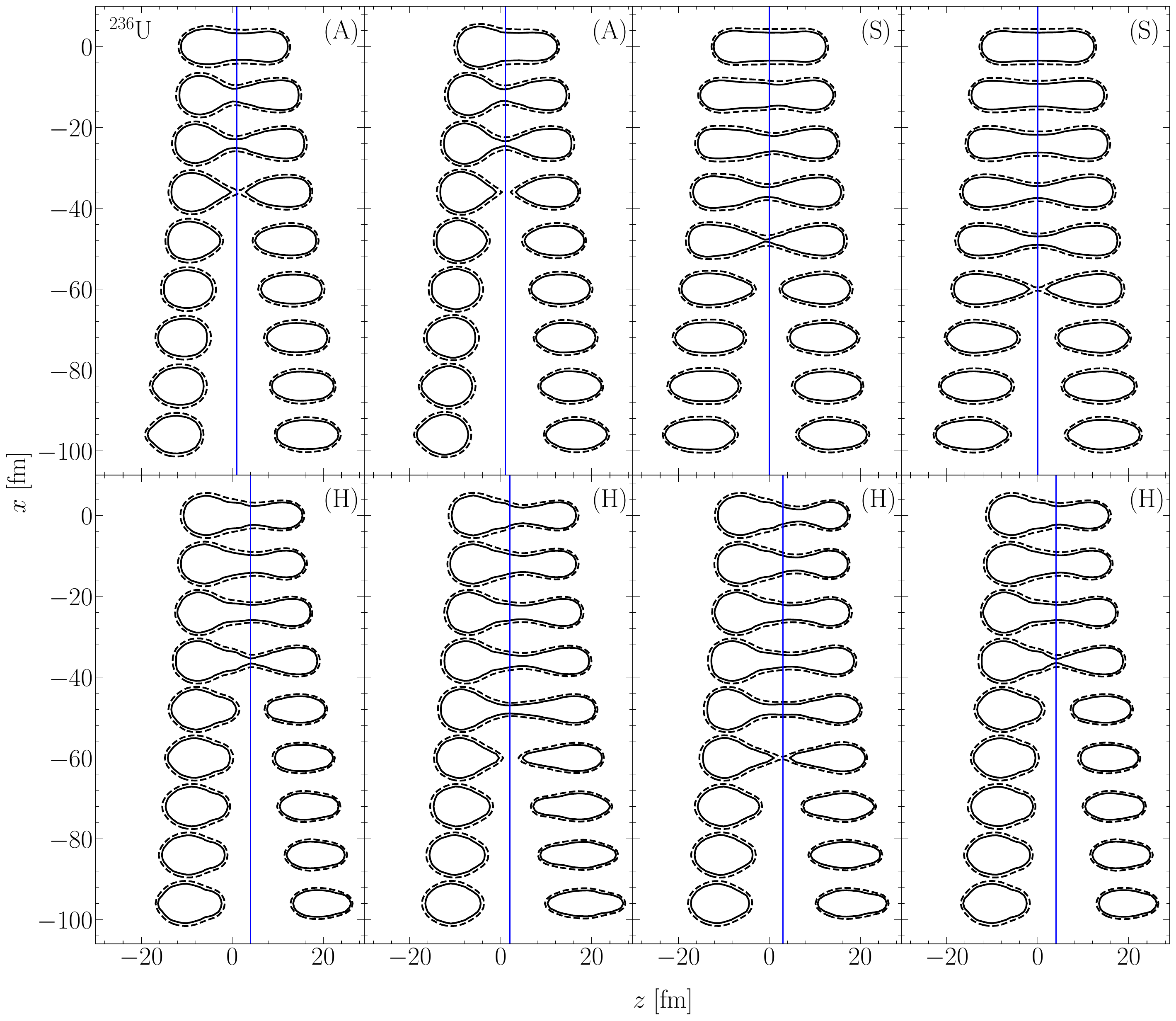}  \caption{ Slices of the neutron and proton number densities for $^{236}$U are shown at fixed separation distances, $d_{\mathrm{sep}} = z^R_{cm} - z^L_{cm}$, where $z^{L/R}_{cm}$ is the center of mass of the system in the left and right volumes respectively, which are separated by a blue line that indicates the position of the neck rupture. The neutron densities are shown with dashed lines and the proton densities are shown with solid lines, both at values of 0.05 $\mathrm{fm}^{-3}$. For each panel the top density configuration is taken at the outer saddle, and every subsequent density is shown at fixed separation distances ranging from 15 fm to 29 fm in 2 fm increments. In the upper left two panels asymmetric fission modes are shown, in the upper right two panels near-symmetric fission modes are shown, and in the bottom four panels highly-asymmetric fission modes are shown. \label{fig:dens2d} } \end{figure*}

Perhaps the starkest distinction between the various fission modes is the shape of the FNS  at scission. In Fig.~\ref{fig:dens2d} a time series of the density profile of the FNS  is shown from saddle to scission and beyond in the case of $^{236}$U for various fission trajectories. The near-symmetric trajectories develop a substantially longer neck than the asymmetric trajectories. For highly-asymmetric trajectories, the situation is more complex, with both shorter and longer necks present. In the language of the Brosa model~\cite{Brosa:1990} it can be stated that the asymmetric trajectories correspond to standard modes, while symmetric trajectories correspond to super long modes, and the highly-asymmetric trajectories can correspond to either.  Speculatively, the presence and absence of the long necks can be explained by the resistance of the primary FFs, those present before scission, to deformation. In asymmetric fission the heavy FF emerges with neutron and proton numbers close to the double magic nucleus $^{132}$Sn. Nuclei close to shell closures tend to be spherical, and this influence could cause the onset of the Rayleigh instability, that ultimately drives the neck ruptures, to appear sooner.  A similar argument can be made for two of the four highly-asymmetric trajectories, where the light FF emerges with neutron number $N \approx 48$, reasonably close to $N = 50$. In these cases, like the heavy FF in asymmetric fission, the light FF is harder to excite, which will discussed more thoroughly in Sec.~\ref{sec:tke}. As argued in~\textcite{Scamps:2018}, there is likely a strong influence of "deformed" magic nuclei on the FF masses and charges. If so, the converse is also true, and there is rich information in the final FF properties, as well as the shape of the FNS  at scission, which is related to the TKE of the FFs, concerning the shell properties of deformed nuclei.  

In the Brosa model, besides the Rayleigh instability, a shift instability is also conjectured to exist. The shift instability allows the "kink" in the neck of the pre-scission FNS  to move freely, provided it is sufficiently shallow, and is the mechanism behind the random neck rupture hypothesis~\cite{Brosa:1990}. In TDDFT, for asymmetric fission, this is clearly not the case from saddle to scission, as the position of the neck rupture is determined at the saddle point and does not move as shown in Fig~\ref{fig:dens2d}, which is consistent with the findings in~\textcite{Abdurrahman:2024}. The situation is more complicated for near-symmetric fission, since the FNS  has no kink at the outer saddle. As the nucleus elongates a kink is formed, and subsequently moves, however its final position does not appear to be random as all trajectories rupture close to the center. This is consistent with the small spread in the final FF charges and masses for these modes, which are listed in Table.~\ref{table:initial}. For highly-asymmetric fission, two possibilities emerge: first, the kink at the outer saddle is fairly close to the location of the neck rupture, shown in the bottom left and right panels in Fig~\ref{fig:dens2d}, second it can move considerably for trajectories with long pre-scission necks, shown in the bottom middle two panels in Fig~\ref{fig:dens2d}. In either scenario, the resulting FFs emerge with significantly more asymmetric masses than the most likely FF configurations, see Table~\ref{table:initial}. This means there is strong evidence that the system at scission retains some memory of the initial mass asymmetry (or equivalently octupole deformation) of the FNS  at the outer saddle point for all modes, and hence the neck rupture is not random. In the present study fluctuations~\cite{Bulgac:2019d} are ignored, which may alter this conclusion in the future.  

\input{Table}



\section{Total Kinetic Energy and Fission Fragment Excitation Energies}\label{sec:tke}

By now it is already well established that the induced fission dynamics from the outer fission barrier to the scission configuration is a highly dissipative process~\cite{Bulgac:2016,Bulgac:2019c,Bulgac:2020,Bender:2020}, 
with, surprisingly, a much longer evolution time than anticipated prior to 2015 in adiabatic microscopic treatments, see Refs.~\cite{Ring:2004,Schunck:2016}. As a result, during the FNS  descent from the top of the outer barrier to the scission configuration the preformed FFs heat-up significantly causing the TKE of the emerging FFs to be much smaller than in adiabatic approaches, and in good agreement with experimental observations. A ``hot'' 
preformed FF deforms much more easily and that is reflected in the increased fraction of emitted neutrons by the HFF at higher neutron incident energies~\cite{Bulgac:2019c,Bulgac:2020}, also in agreement with observations.     

At the neck rupture, when no more nucleons are transferred between particles, the total kinetic energy is determined almost exclusively by the following ingredients, the charges of the heavy and light fragments ($Z_{H/L}$), the separation distance between the center of masses of the fragments ($d_{\mathrm{rupt}}$), and the collective kinetic energy at the time of the rupture ($K_{\mathrm{rupt}}$),
\begin{equation}
\label{eqn:tke}
TKE \approx \frac{e^2 Z_L Z_H}{d_{\mathrm{rupt}}} + K_{\mathrm{rupt}},
\end{equation}
with minor corrections due to FF deformations and particle emission at scission.  

\begin{figure} \includegraphics[width=1.0\columnwidth]{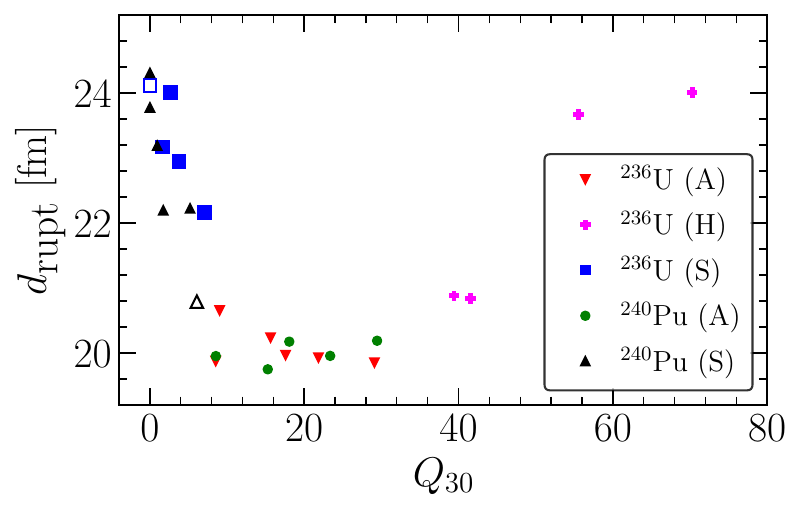}  \caption{ \label{fig:scidist}  The separation distance between the proto-fission fragments at scission is shown as a function of the FNS  initial octuple moment at the outer saddle for $^{236}$U and $^{240}$Pu.  The labels (A), (S), and (H) refer to asymmetric fission modes, near-symmetric fission modes, and highly-asymmetric fission modes respectively. The open black triangle represents an intermediate trajectory in which the FNS  initially had a small octupole deformation, but fissions  asymmetrically. The open blue square represents an exactly mass symmetric fission trajectory.}  \end{figure}

As shown in Fig.~\ref{fig:scidist} the FNS s in both near-symmetric and highly-asymmetric fission have a consistently larger separation distance when the neck ruptures, relative to asymmetric fission. The value of this distance is strongly correlated with the initial mass asymmetry of the FNS  at the outer saddle, which determines which "valley" the system tends to as it descends to scission. The greater rupture separation distance is the main reason near-symmetric fission FFs emerge with much lower total kinetic energy ($\sim 25$ MeV lower) relative to typical asymmetric FFs. This is clear since the difference in the collective kinetic energy at the rupture for these two modes, recorded in Table.~\ref{table:initial},  can only account for $\sim 5$ MeV, meanwhile the FF charge product $Z_L Z_H$ is larger for near-symmetric fission and thus acts to increase the TKE, opposite to the observed trend. In contrast, in highly-asymmetric fission, all elements conspire together to produce a lower TKE. On average, these modes have the smallest collective kinetic energy at the rupture and the smallest FF charge product of any fission mode. They also rupture at larger separation distances compared to asymmetric fission, albeit with much larger variance, as shown in Table.~\ref{table:initial}.   

\begin{figure} \includegraphics[width=1.0\columnwidth]{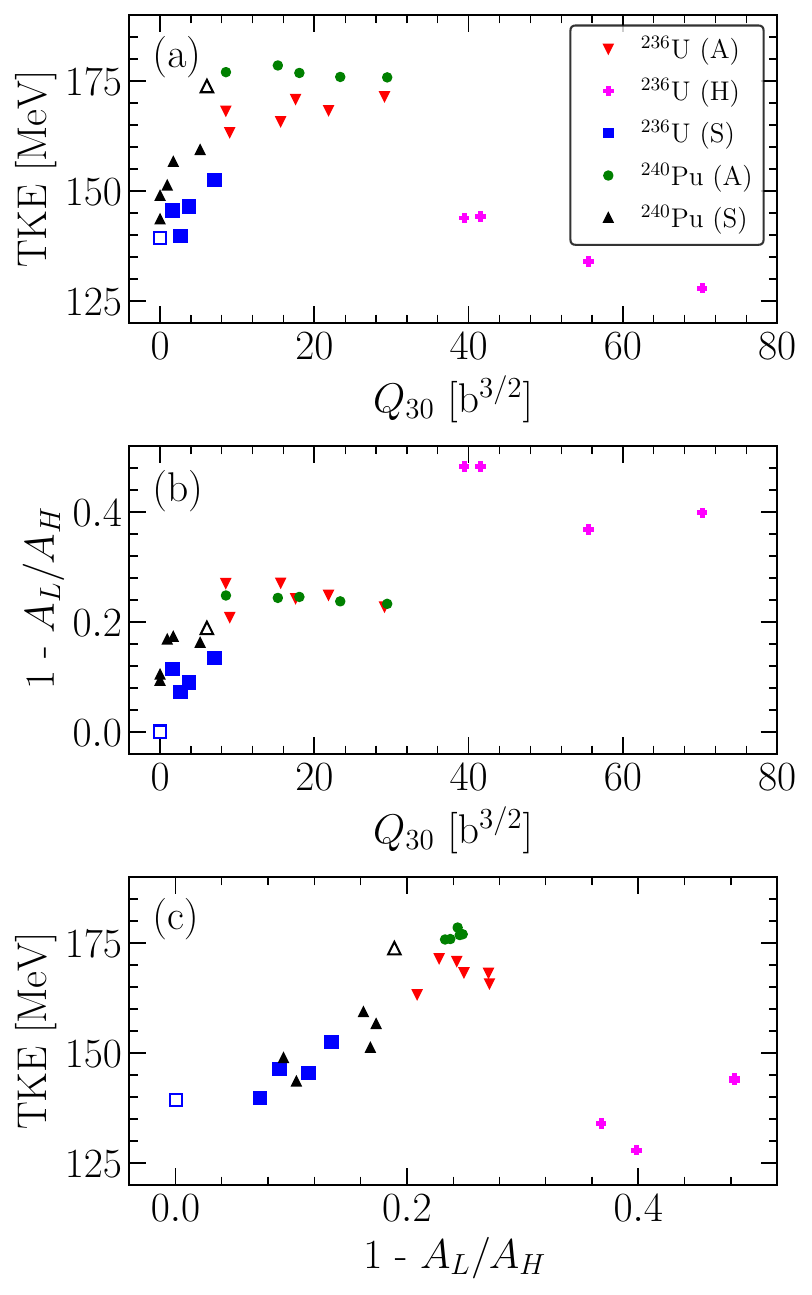}  
\caption{Panel (a) shows the TKE vs. the initial octupole moment of the FNS s $^{236}$U and $^{240}$Pu at the outer saddle for various fission modes defined previously (see the caption of Fig.~\ref{fig:scidist} for details). Panel (b) shows the mass asymmetry of the FFs vs. the initial octupole moment of the FNS  for the same reactions.  Panel (c) shows the TKE vs the final mass asymmetry of the FFs for the same reactions. \label{fig:TKE}}  \end{figure}

The lower TKE associated with near-symmetric fission trajectories could potentially shed light on a phenomenon known from experiment, but still lacking a satisfactory microscopic explanation, the fact that the total TKE decreases as the excitation energy of FNS  increases~\cite{Madland:2006}. At lower FNS excitation energies, such as the case of fission with thermal neutrons, near-symmetric fission trajectories are suppressed due to the presence of the second barrier along $Q_{30}=0$, see Fig.~\ref{fig:trajs}. It is believed that along the lowest-energy path from the ground state to scission the FNS  starts axial, undergoes a triaxial shape transition at the first barrier, and undergoes a reflection asymmetric shape transition at the second barrier~\cite{Ryssens:2015}, before reaching the ridge of the asymmetric valley. However, when the compound's excitation energy is higher, the nucleus is less sensitive to the structure of the (lowest-energy) PES, and becomes increasingly likely to take the path directly over the higher symmetric barrier, thereby increasing the population of near-symmetric fission modes, and, consequentially, lowering the average value of the TKE, see Fig.~\ref{fig:TKE}. Ideally, the framework would be able to assign weights to the initial configurations, allowing this claim to be investigated more directly. Unfortunately, to the authors' knowledge, there is no rigorous and well justified microscopic prescription for assigning these weights at the present.  

Fig.~\ref{fig:TKE} also makes clear the strong correlation between the TKE, the initial mass asymmetry of the system, characterized by the compound's octupole deformation, and the final mass asymmetry of the system, characterized by the ratio of the FF masses, $R \equiv 1 - A_L/A_H$. The relationship between the FF masses and the TKE is similar to what was observed in~\cite{Bjelvcic:2026}. This reemphasizes that, for each class of fission trajectory, the system retains memory of the FNS s properties at the outer saddle throughout its evolution towards scission. Whether the same is true within each class remains an open question.  

\begin{figure*} \includegraphics[width=2.0\columnwidth]{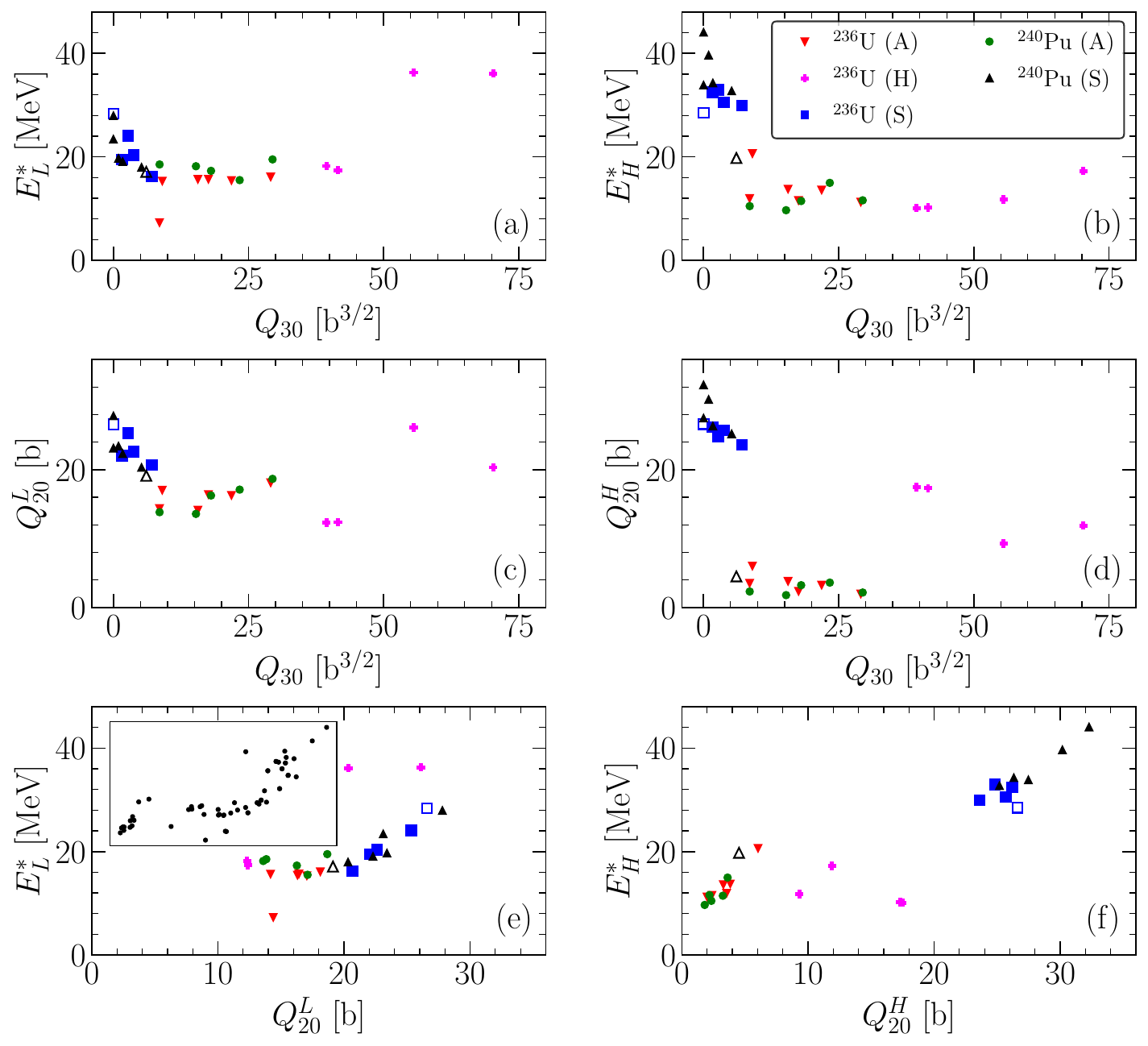}  
\caption{Panels (a) and (b) show the excitation energies of the light and heavy FFs vs. the initial octupole moment of the FNS  at the outer saddle. Panels (c) and (d) show the quadrupole moment of light and heavy FFs vs. the initial octupole moment of the FNS  at the outer saddle. Panels (e) and (f) show the excitation energies of the light and heavy FFs vs. the quadrupole moment of the FFs. The various fission modes (colors and markers) are described in the caption of Fig.~\ref{fig:scidist}. The inset shows equivalent information to the bottom two panels, without distinguishing between fission modes or fragment types. \label{fig:Estarall}}  \end{figure*}

The lower TKE present in both near-symmetric and highly-asymmetric fission, relative to asymmetric fission, is compensated by the excitation energies of the FFs. The top row of panels in Fig.~\ref{fig:Estarall} reveals how this energy is shared between the FFs. In near-symmetric fission, the HFF obtains most of the additional excitation energy, although the LFF is slightly more excited as well. In highly-asymmetric fission the LFF gets significantly more excited for the two trajectories with longer necks at scission, while for the other two cases the FF excitation energies are comparable to asymmetric fission. 

The second row of panels reveal that the HFF gets significantly more deformed in near-symmetric fission, relative to asymmetric fission. This is likely because the HFF neutron and proton numbers greatly differ from the double magic nucleus $^{132}$Sn or the deformed magic nucleus discussed in~\textcite{Scamps:2018}. Being away from the shell closures the HFF is easier to deform. In highly-asymmetric fission the HFF is more deformed than in the case of asymmetric fission and less deformed than in the case of near-symmetric fission.  For the LFF the values fluctuate wildly. 

The relationship between the FF deformations, specifically their quadrupole moments, and their excitation energies is shown in the bottom row of panels. In near-symmetric fission, most of the additional excitation energy goes into the HFF, and specifically into deforming the HFF. The additional excitation energy gained by the LFF is also strongly correlated with its quadrupole deformation. In fact, in near-symmetric fission, there is almost an exact linear correlation between the FF excitation energy and its quadrupole deformation. In contrast, asymmetric fission exhibits almost no correlation between the deformation of the FFs and their excitation energies. 
For highly-asymmetric fission, the situation is more complicated, and likely requires more trajectories to define a clear trend. 

In the inset of Fig.~\ref{fig:Estarall} all FF excitation energies are shown as a function of their quadrupole moment without distinguishing between classes or if the fragment is heavy or light. This highlights are more general trend: two distinct regions, separated by a critical deformation of $Q_{20} \approx 20$ b. At values below this deformation, a FF's excitation energy and quadrupole moment are almost entirely uncorrelated. Above this value, there is a strong linear correlation, and the FF requires energy to deform. The converse suggests that the final excitation energy of the FFs, if sufficiently high, can be used to probe the deformation of the FFs before the emission of prompt neutrons. 

These results further reveal information concerning the dependence of the FF excitation energy sharing as a function of the incident neutron's energy. At low neutron energies the LFF and HFF will have comparable excitation energies, as asymmetric fission is the dominant mode. At higher neutron energies, near-symmetric fission becomes common, and the additional excitation energy goes primarily into the HFF. This trend, albeit substantially weaker, was also observed in previous studies only considering excited asymmetric fission modes~\cite{Bulgac:2021}. 

Note, TDDFT, in its current implementation, underestimates the TKE of the rarer fission modes, and, consequentially, overestimates their FF excitation energies when compared to experiment. This was also seen in other studies~\cite{Bjelvcic:2026}. This is likely due to missing beyond mean field correlations, such as the inclusion of proton neutron collision terms. Despite this, although TDDFT is not quantitatively exact at the moment, many of the predicted qualitative trends are still expected to hold.  

\section{Neck Rupture}\label{sec:neckr}

\begin{figure} \includegraphics[width=1.0\columnwidth]{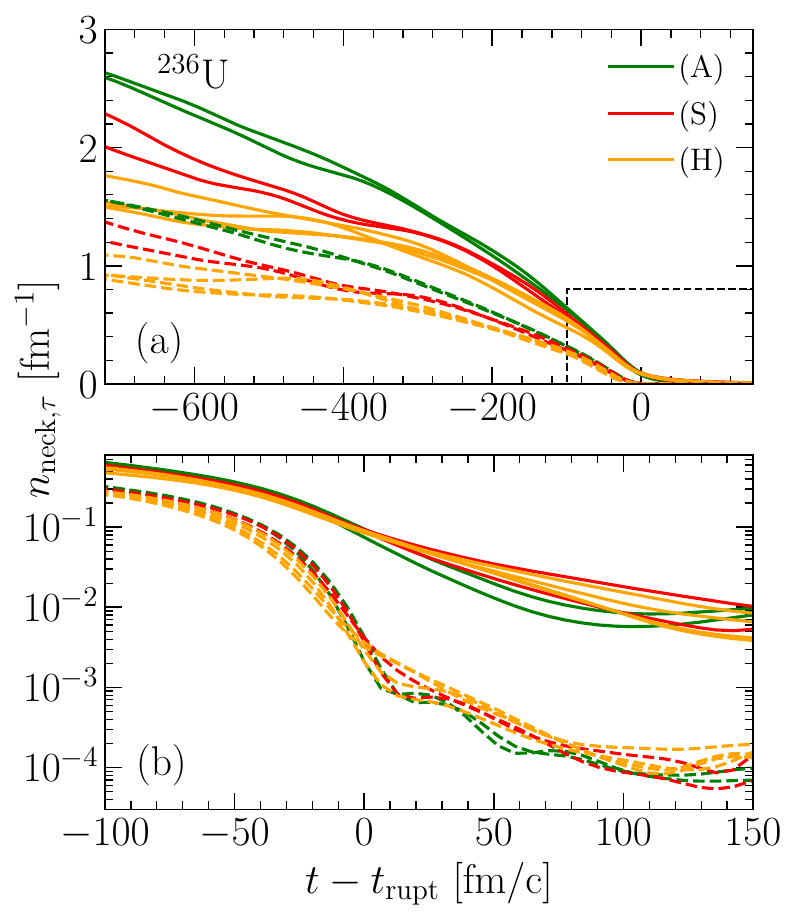}
\caption{ In panel (a) the integrated neck density is shown as a function of time, with respect to the time of the neck rupture, for the $^{236}$U trajectories shown in Fig.~\ref{fig:dens2d}. The neck density is defined by $n_{\mathrm{neck},\tau} = \mathrm{min} \int\big[n_\tau(x,y,z)\big ]dx dy$, where $z$ is the fission axis and $\tau =n,p$. The solid/dashed lines denote the neutron/proton necks respectively. As previously, the labels (A), (S), and (H) refer to asymmetric fission modes, near-symmetric fission modes, and highly-asymmetric fission modes respectively. In panel (b) the integrated neck density is shown in log scale for a more restricted interval of time. The limits of panel (b) are outlined by the rectangle with black dashed lines in panel (a).  \label{fig:neck}} \end{figure}

This section investigates the neck rupture, the most non-equilibrium stage of fission, for various fission modes. Fig.~\ref{fig:neck} shows how the neck decays for a few $^{236}$U fission trajectories: the same ones presented in Fig.~\ref{fig:dens2d}. As seen in panel (a), both the proton and neutron necks decay slower for near-symmetric and highly-asymmetric fission, relative to asymmetric fission, at times away from the neck rupture.  At times closer to the rupture, shown in panel (b), the situation is more complicated. For asymmetric fission, the proton and neutron necks decay exponentially $n_{\mathrm{neck},\tau} \sim e^{\frac{-t}{\tau}}$. For near-symmetric fission, and the two cases of highly-asymmetric fission with longer necks, a second exponential decay is present at slightly later times for the neutron neck densities. In a few instances, this second exponential decay is also present for the proton neck densities. 


\begin{figure} \includegraphics[width=1.0\columnwidth]{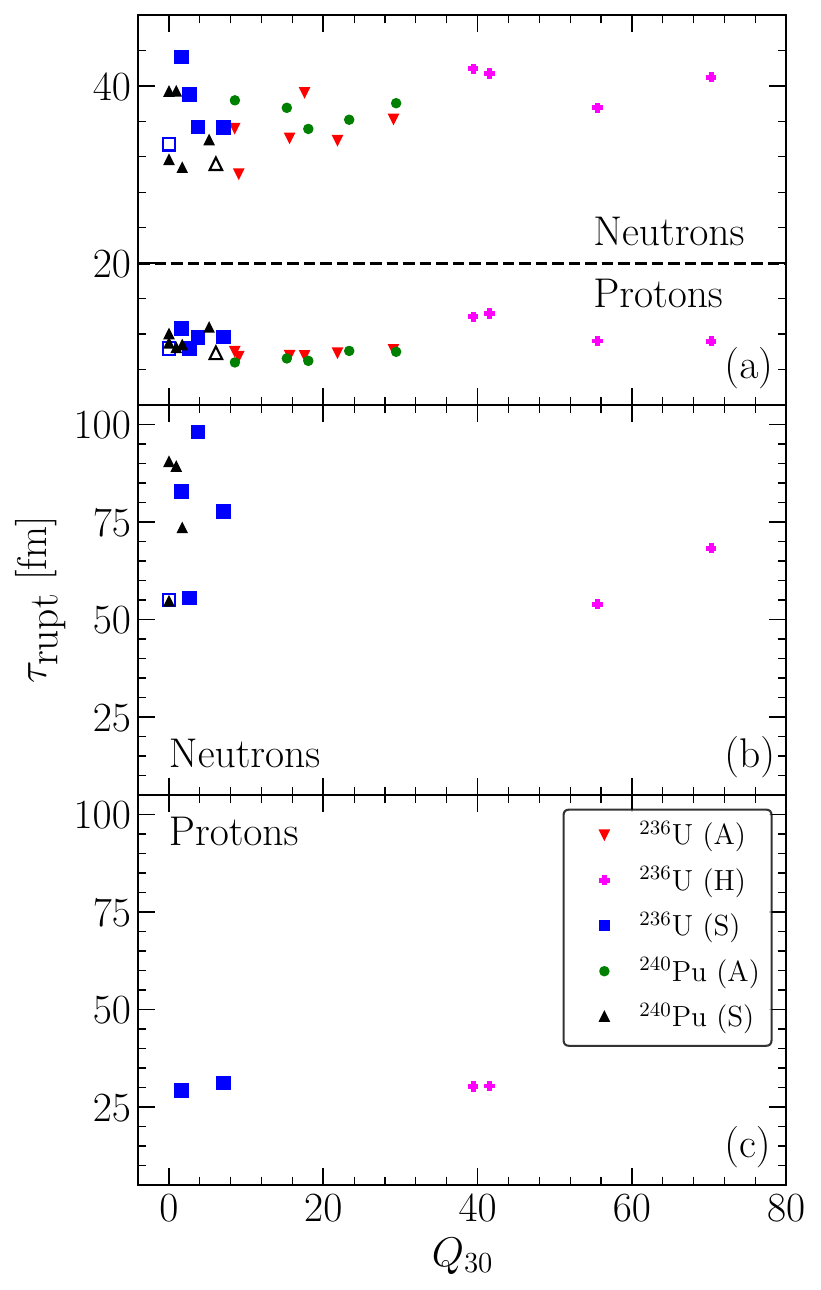}
\caption{ Panel (a) shows the rupture time for the neutron and proton necks as a function of the compound's initial octupole moment at the outer saddle. The proton and neutron times are divided by a dashed black line. The various fission modes (colors and markers) are described in the caption of Fig.~\ref{fig:scidist}. Panel (b) shows the rupture time for certain neutron trajectories, where a second exponential decay of the neck is present. Panel (c) shows the same for protons. In all cases the rupture time was extracted from a fit to a decaying exponential $n_{\mathrm{neck}} \sim \exp\left (-\tfrac{t}{\tau}\right )$.  In some instances two distinct timescales emerged, requiring two separate fits. \label{fig:taurupt}} \end{figure}

The rupture time $\tau$ is provided, and defined, in Fig.~\ref{fig:taurupt}. As was seen in~\textcite{Abdurrahman:2024}, the first exponential decay (the top panel of Fig.~\ref{fig:taurupt}) is universal, with the rupture time roughly equal for all trajectories. Furthermore, the proton neck always ruptures faster than the neutron neck. In panel (b) another rupture time $\tau$ is shown, characterizing a second exponential decay present for near-symmetric fission and the two highly-asymmetric fission trajectories with long necks at scission. This second time scale is present because the Coulomb repulsion between the FF is weaker at the rupture for near-symmetric fission. This causes the FFs to move away from each other at a slower pace. Simultaneously the proton rupture is typically unaffected, as at late stages of the rupture, the neck is almost always exclusively sustained by neutrons. Interestingly, there are a few exception to this rule, as the two highly-asymmetric fission trajectories with shorter necks exhibit a second decay of the proton neck, shown in panel (c). This is also the case for two near-symmetric fission trajectories of $^{236}$U. 

Another question can now be posed, how do differences in the neck rupture between the various modes affect the emission of scission neutrons (SNs): neutrons emitted during the neck rupture. Such neutrons were conjectured to exist as early as 1939 by~\textcite{Anderson:1939} and~\textcite{Bohr:1939}, and recently shown to be present in a fully microscopic model~\cite{Abdurrahman:2024}. To address this, one simulation of a near-symmetric fission trajectory was performed in a $48\times48\times96$ point lattice with 1 fm spacing. Ideally, highly-asymmetric fission would have included as well.  However, the computational cost of TDDFT fission calculations in such large lattice is so great that only one trajectory could be included in the present study. 

\begin{figure} \includegraphics[width=1.0\columnwidth]{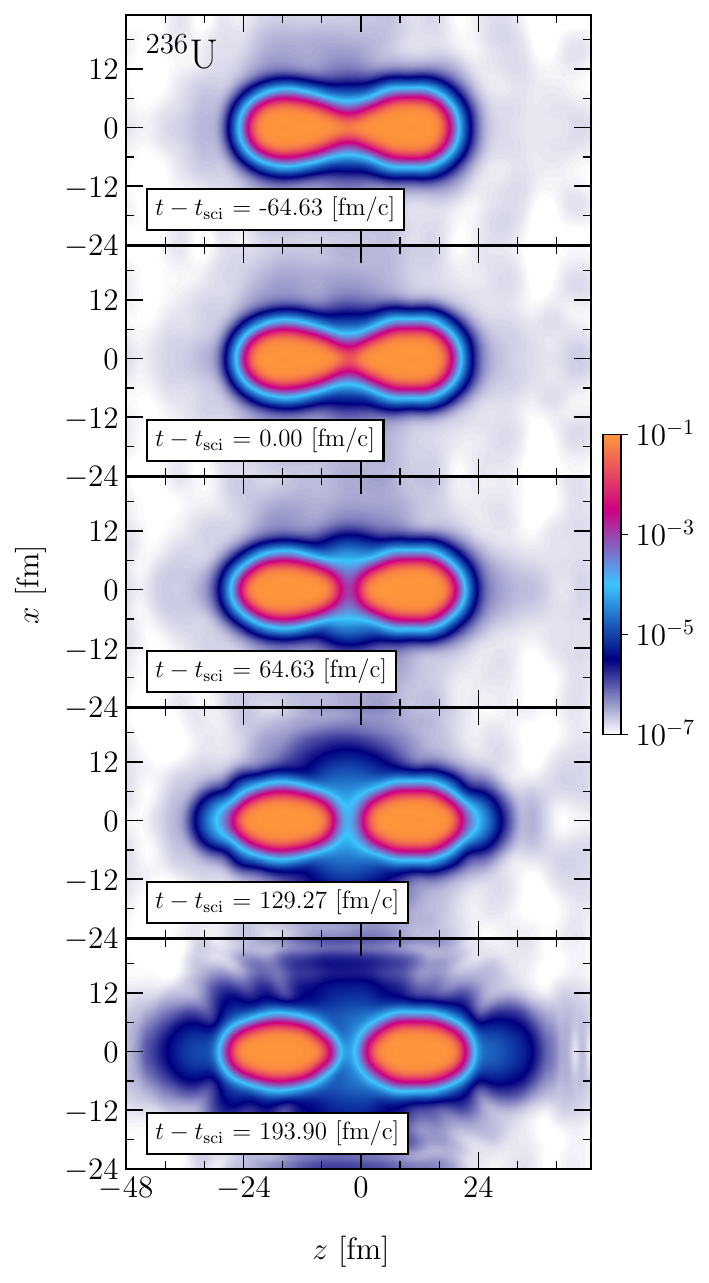}  \caption{ Time series of slices of the neutron number density in $\mathrm{fm}^{-3}$ for a near-symmetric fission trajectory. The color scheme is the same as was used in~\cite{Abdurrahman:2024}. \label{fig:scipar}  }  \end{figure}

\begin{figure} \includegraphics[width=1.0\columnwidth]{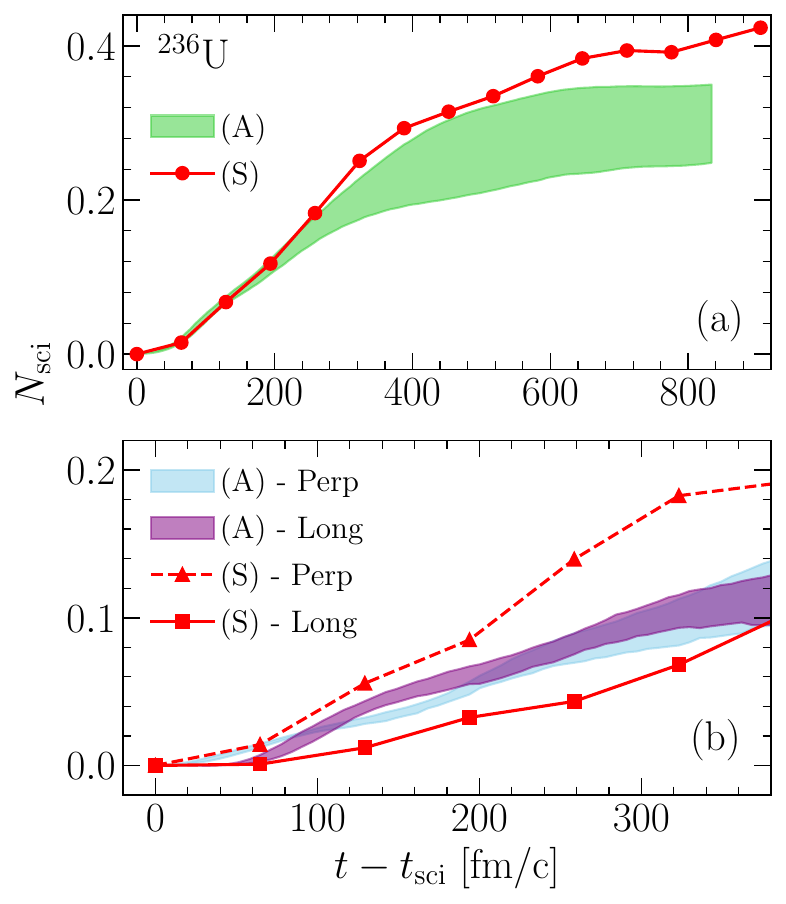}
\caption{ Panel (a) shows the number of scission neutrons released as a function of time for $^{236}$U. The red line represents a near-symmetric fission trajectory, while the green region represents the mean $\pm$ one standard deviation for typical asymmetric fission modes (taken from~\textcite{Abdurrahman:2024}). Panel (b) breaks down the SN emission into neutrons emitted along (Long), and perpendicular (Perp) to, the fission axis. In the lower panel, times are only considered until the longitudinal clouds hit the boundary of the simulation box. 
\label{fig:nsci}} \end{figure}

Fig.~\ref{fig:scipar} shows the neutron density profile for this run, exhibits the same universal signature observed for asymmetric fission~\cite{Abdurrahman:2024}: a cloud of neutrons emitted perpendicular to the fission axis, followed by clouds emitted in front of the FFs. This was also recently confirmed in~\cite{Bjelvcic:2026b} using a different EDF. The total SN emission, shown in panel (a) of Fig.~\ref{fig:nsci}, is also comparable, with only slightly more neutrons emitted in near-symmetric fission. However, the character of the emission is different. In panel (b) of Fig.~\ref{fig:nsci}, the SN emission is decomposed into neutrons emitted in front of the FFs vs neutrons emitted perpendicular to the fission axis. For asymmetric fission both components are comparable, while for near-symmetric fission significantly more neutrons are emitted perpendicular to the fission axis than parallel to it. This stems from the fact that the neck is longer and wider in near-symmetric fission, allowing for more neutrons to be emitted between the FFs during the rupture. Simultaneously, the rupture is also a slower process for  near-symmetric fission, meaning the force produced from the retraction of the FF tails during the rupture is weaker, causing less longitudinal neutrons to be ejected. This effect, known as the catapult mechanism~\cite{Madler:1985}, is believed to be responsible for the emission of SNs in front of the FFs. 

\section{Conclusion}\label{sec:con}

In summary, this study investigated the neutron-induced fission of $^{235}$U and $^{239}$Pu from saddle to scission, in TDSLDA, for a few classes of trajectories distinguished by the value of the octupole deformation at the outer saddle. It was found that near-symmetric trajectories exhibit  striking differences when compared to their asymmetric counterparts, all of which stem from the dramatically elongated neck of the FNS  prior to scission. The TKE is substantially lower for these modes, as the elongated neck leads to a larger separation distance between the proto-fragments at the neck rupture, thereby lowering the Coulomb interaction energy between the FFs. The increased population of these modes, as the energy of the incident neutron increases, could potentially explain the observed experimental trend that the average TKE in fission decreases as a function of the incident neutron's energy. The lower TKE correlates to higher total excitation energy, the majority of which goes into the HFF, which emerges highly deformed. Near-symmetric modes also take longer to fission and exhibit a far more complex evolution from saddle to scission.  

In contrast, highly-asymmetric modes contain more variety, forming both longer and shorter necks. When longer necks are formed the LFF emerges substantially more excited than the HFF. In both cases the TKE decreases, although the reasons differ.

The neck rupture was investigated in all cases, with two distinct phases observed in near-symmetric fission. The first exponential decay of the neck occurs at a universal time-scale for all fission modes, while the second is slower and only occurs in certain instances. This slower neck rupture influences the SN signal for near-symmetric fission. The same universal signal was still observed, neutron clouds emitted perpendicular and parallel to the fission axis, as well as roughly the same number of SN emitted as in asymmetric fission~\cite{Abdurrahman:2024}.  However, for near-symmetric fission significantly more neutrons were observed perpendicular to the fission axis than parallel to it. In comparison, for typical asymmetric fission, the numbers are roughly equal. 

Although not investigated here, it ought to be mentioned that the FF spin properties could be vastly different between the various fission modes. The magnitude of the intrinsic FF spins will likely be different for rarer fission modes compared to the asymmetric fission mode~\cite{Bulgac:2021}, which is expected from both experiment~\cite{Wilson:2021} and adiabatic approaches~\cite{Marevic:2021}. The correlations between the FF total angular momenta vectors could also potentially differ greatly~\cite{Bulgac:2022b,Scamps:2023a}. In the future, such correlations should be extracted not only within TDSLDA, but experimentally as well.  

This study focuses the influence of the various exit channels, at the outer saddle, on fission observables. Due to conservation of energy and the fact that the FNS  must deform smoothly, the system will pass through these configurations, at least when axial symmetry is assumed and the multipole shape expansion is truncated at third order. When the system fissions through rarer modes, many trends observed in asymmetric fission no longer hold true, highlighting how complex saddle to scission dynamics really is. The key take away is, in order to fully understand fission, all such modes (and more) should be throughly investigated at the microscopic level. 

\section*{Acknowledgments}\label{sec:ack}

I.A. and I.S. were supported by the U.S.
Department of Energy through the Los Alamos National
Laboratory. The Los Alamos National Laboratory is operated
by Triad National Security, LLC, for the National Nuclear
Security Administration of the U.S. Department of
Energy Contract No. 89233218CNA000001. I.A. and I.S. gratefully acknowledge partial support and computational
resources provided by the Advanced Simulation and
Computing (ASC) Program. M.K. was supported by NNSA cooperative Agreement DE-NA0003841. A.B. was supported by the Office of Science, Grant No. DE-FG02-97ER41014  
and partially by NNSA cooperative Agreement DE-NA0003841. This research used resources of the Oak Ridge Leadership Computing 
Facility, which is a U.S. DOE Office of
Science User Facility supported under Contract No. DE-AC05-00OR22725.


\providecommand{\selectlanguage}[1]{}
\renewcommand{\selectlanguage}[1]{}

\bibliography{local_fission}

\clearpage

\end{document}

%% file: Table.tex
\begin{table*}[!htb]
\centering
\caption{ The first column defines the compound system and type of fission mode considered, where (A) represents asymmetric fission, (S) represents near-symmetric fission, (E) represents exactly-symmetric fission, (H) represents highly-asymmetric fission, and (I) represents an intermediate trajectory, see the magenta curve in Fig.~\ref{fig:trajs} for more details. It also contains the number of fission trajectories considered for each mode. Across the remaining columns, and two rows, the mean value and standard deviations of the following quantities are recorded: the initial quadrupole and octupole deformations of the compound system, the charges of FFs, the masses of the FFs, the final quadrupole deformations of the FFs, the final octupole deformations of the FFs, the excitation energies of the FFs, the separation distances and times at the neck rupture, and the TKE and collective kinetic energy at the rupture. The neck rupture is defined as the point in time when the integrated density $n_{\mathrm{neck}} = \mathrm{min}\int\big[n_n(x,y,z) + n_p(x,y,z)\big ]dx dy$ first falls below 0.1 fm$^{-1}$.  The collective kinetic energy is computed directly from Eq.~\ref{eqn:tke}. }
\begin{tabular}{rrrrrrrrr}
\hline \hline
Nucl. & $Q_{20}$ [b] & $Z_H$ & $A_H$ & $Q_{20}^{H}$ [b]& $Q_{30}^{H}$ [b$^{3/2}$]& $E^*_H$ [MeV]& $d_\textrm{rupt}$ [fm] & $TKE$ [MeV] \\
No. Runs & $Q_{30}$ [b$^{3/2}$]& $Z_L$ & $A_L$ & $Q_{20}^{L}$ [b] & $Q_{30}^{L}$ [b$^{3/2}$] & $E^*_L$ [MeV] & $t_\textrm{rupt}$ [fm/c]& $K_\text{rupt}$ [MeV] \\ \hline\hline

$^{236}$U (A) & 161.32 (18.08) & 51.99 (0.53) & 134.51 (1.70) & 3.51 (1.30) & 0.33 (0.07) & 13.80 (3.21) & 20.09 (0.28) & 168.02 (2.81)  \\  
6             & 16.95 (7.16) & 40.01 (0.53) & 101.49 (1.70) & 16.08 (1.40) & 0.12 (0.06) & 14.26 (3.13) & 2232 (731) & 18.91 (1.19) \\ \hline

$^{236}$U (H) & 265.81 (31.88) & 58.07 (1.85) & 150.77 (4.90) & 14.00 (3.52) & 0.61 (0.56) & 12.32 (2.92) & 22.35 (1.49) & 137.49 (6.89)  \\   
4             & 51.70 (12.39) & 33.93 (85) & 85.23 (4.90) & 17.79 (5.80) & 0.41 (0.34) & 27.01 (9.21) & 2177 (332) & 10.36 (1.32) \\ \hline

$^{236}$U (S) & 198.44 (0.32) & 48.39 (0.63) & 124.43 (1.56) & 25.08 (0.99) & 0.31 (0.11) & 31.47 (1.27) & 23.07 (0.66) & 144.70 (4.84) \\   
4& 3.79 (2.05) & 43.59 (0.61) & 111.56 (1.54) & 22.66 (1.70) & 0.45 (0.06) & 20.02 (2.80) & 5373 (680) & 14.30 (0.91) \\ \hline

$^{236}$U (E) & 199.43 & 46.01 & 118.02 & 26.62 & 0.36 & 28.49 & 24.12 & 139.32 \\   
1& -5.56$\times10^{-8}$ & 45.99 & 117.98 & 26.58 & 0.36 & 28.38 & 7116.10 & 12.99 \\ \hline\hline

$^{240}$Pu (A) & 160.80 (20.77) & 53.09 (0.21) & 136.49 (0.43) & 2.64 (0.68) & 0.28 (0.10) & 11.63 (1.81) & 20.00 (0.16) & 176.82 (0.98) \\   
5& 18.94 (7.10) & 40.91 (0.21) & 103.51 (0.43) & 15.90 (1.94) & 0.25 (0.06) & 17.80 (1.35) & 1569 (183) & 20.47 (0.91) \\ \hline

$^{240}$Pu (S) & 197.98 (2.44) & 50.34 (0.93) & 129.09 (2.34) & 28.29 (2.60) & 0.24 (0.14) & 36.91 (4.31) & 23.14 (0.84) & 151.94 (5.60)  \\   
5& 1.57 (1.93) & 43.64 (0.97) & 110.88 (2.38) & 23.39 (2.46) & 0.54 (0.27) & 21.60 (3.66) & 4751 (1671) & 15.14 (1.47) \\ \hline

$^{240}$Pu (I) & 188.78 & 51.30 & 132.53 & 4.53 & 0.30 & 20.79 &  24.5 & 173.85 \\  
1& 6.08 & 42.70 & 107.46 & 19.13 &  0.25 & 17.10 & 1364 & 22.14 \\ 
\hline \hline
\end{tabular}
\label{table:initial}
\end{table*}